\documentclass[twocolumn,showpacs,preprintnumbers,amsmath,amssymb]{revtex4} 
                                                                                
\usepackage{graphicx}
\usepackage{dcolumn}
\usepackage{bm}
\begin{document}
                                                                                
\preprint{CUPhys/12/2006}
                                                                                
\title{Floating Phase in 2D ANNNI Model}
                                                                                
\author{Anjan Kumar Chandra }
\author{Subinay Dasgupta}%
\affiliation{%
Department of Physics, University of Calcutta, 92 Acharya Prafulla Chandra Road, Calcutta 700009, India.\\
}%
                                                                                
\date{\today}
                                                                                
\begin{abstract}
We investigate whether the floating phase
(where the correlation length is infinite and the spin-spin
correlation decays algebraically with distance) exists in the
temperature($T$) - frustration parameter ($\kappa$) phase diagram of 2D ANNNI
model.
To identify this phase, we look for the region where (i) finite size effect
is prominent and (ii) some relevant physical quantity changes somewhat
sharply and this change becomes sharper as the system size increases.
For $\kappa < 0.5 $, the low temperature phase is ferromagnetic and we
study energy and magnetization. For $\kappa > 0.5 $, the low temperature
phase is antiphase and we study energy, layer magnetization, length of domain
walls running along the direction of frustration, number of domain-intercepts
that are of length 2 along the direction of frustration, and the number
of domain walls that do not touch the upper and/or lower boundary. In 
agreement with some previous studies, our final conclusion is that, the 
floating phase exists, if at all, only along a line.
\end{abstract}
                                                                                
\pacs{05.70.Jk, 05.10.Ln, 64.60.Fr}
                                                                                
\def\be{\begin{equation}}
\def\ee{\end{equation}}
\maketitle

\section{Introduction} 

The two-dimensional Axial Next-Nearest Neighbor Ising (ANNNI) model 
(${\rm spin}=\frac{1}{2}$) is a square lattice 
Ising model with nearest neighbor ferromagnetic interaction along both the
axial directions and second neighbor anti-ferromagnetic interaction along 
one axial direction. The Hamiltonian is,
\begin{equation}
{\mathcal H} = - J \sum_{x,y} s_{x,y} [s_{x+1,y}+s_{x,y+1}-\kappa s_{x+2,y}]
\end{equation}
where the sites ($x$,$y$) runs over a square lattice, the spins $s_{x,y}$ are
$\pm 1$, $J$ is the nearest-neighbor interaction strength and $\kappa$ is a
parameter of the model. For positive values of $\kappa$ the second-neighbor
interaction introduces a frustration.
This is one of the simplest frustrated classical Ising model
with a tunable frustration and has been studied over a long time 
(\cite{selke,yeomans,liebman} for review). The most widely studied aspect of 
this model is the phase diagram in the $T-\kappa$ phase space, where $T$ 
stands for temperature. It is easy to prove analytically that 
\cite{selke,yeomans,liebman} at 
zero temperature, the system is in a ferromagnetic state for $\kappa < 0.5$,
and in antiphase ($++--++-- \cdots$ along $x$ direction and all like spins 
along $y$ direction) for $\kappa > 0.5$ with a ``multiphase'' state at
$\kappa = 0.5$. (The multiphase state comprises of all possible configurations
that have no domain of length 1 along the $x$ direction.)

\begin{figure}
\noindent \includegraphics[width= 8cm, angle = 0]{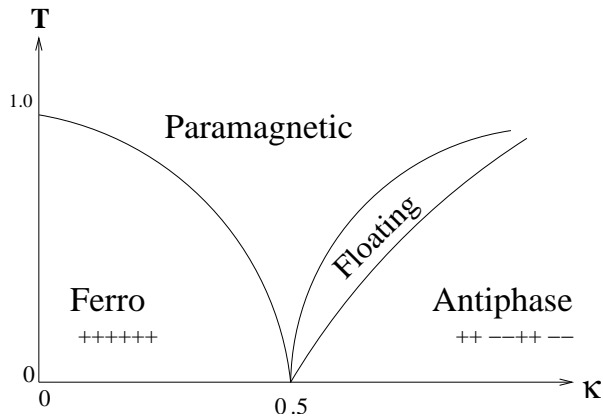}
\caption{\label{fig:ani}Schematic phase diagram of the two-dimensional ANNNI 
model according to previous studies. 
}
\end{figure}

From Monte Carlo simulations and approximate analytic calculations, some
early studies had proposed a phase diagram (Fig.~\ref{fig:ani}) consisting 
of a ferromagnetic phase (for $\kappa < 0.5$) and antiphase (for 
$\kappa > 0.5$) 
at low temperature along with a paramagnetic phase at high temperature (for 
all $\kappa$ values). The crucial point is, between the ordered and the 
disordered phases for $\kappa > 0.5$, 
there may be a so-called ``floating'' phase characterized by (i) a spin-spin 
correlation that decays algebraically with distance and (ii) an incommensurate,
continuously varying modulation. 
An approximate analytic treatment by Villain and Bak \cite{bak} predicted that 
(for $\kappa > 0.5$) as temperature 
increases from zero, there is a second order Pokrovsky-Talapov type 
commensurate-incommensurate phase transition from antiphase to the floating 
phase followed by a Kosterlitz-Thouless transition from the floating to the
paramagnetic phase. This leads to a phase diagram shown schematically in
Fig. 1. Several computational studies (see \cite{selke,jap})
also confirmed such a phase diagram. Later, Shirahata and 
Nakamura \cite{jap} have measured the dynamical exponent by studying the 
non-equilibrium relaxation of order parameter at $\kappa =$ 0.6 and 0.8 and 
concluded that the floating phase exists, if at all, over a narrow 
temperature range only. The central problem in this study is that the 
identification of order parameter is ambiguous for $\kappa > 0.5 $, in the 
sense that it is difficult to identify a physical observable that relaxes
algebraically at the critical temperature. One may note that the antiphase
magnetization
\[ M_{<2>} = \sum_{x,y} s^{<2>}_{x,y}s_{x,y} \]
does not satisfy this criterion \cite{jap}. (Here, $s_{x,y}^{<2>}$ is the
spin distribution for perfect aniphase distribution.)
Recently, a density matrix renormalization group analysis \cite{jap2} has also 
excluded the presence of any incommensurate phase over an extended region.

The 2D ANNNI model is related to the transverse ANNNI chain by Suzuki-Trotter 
transformation \cite{suzuki,ariz,bkc_book,arnab}. This quantum Ising model 
has also been studied widely. Numerical and approximate analytic studies
\cite{bkc_book,rieger,amit} had predicted that the floating phase exists in 
the transverse ANNNI chain over a region, as shown in Fig. 1. However, 
recently we have shown \cite{CD2} that for this model, the floating phase 
exists over a wide region extending from $\kappa < 0.5$ to $\kappa > 0.5$.
There is thus a controversy about the existence of algebraically decaying
phase in 2D ANNNI model.

In this article, we shall present Monte Carlo simulation of 2D ANNNI model
with a view to locating the floating phase, if any.
For $\kappa < 0.5 $, one can easily identify the magnetization as the order 
parameter and for this case we have therefore, measured 
\begin{enumerate}
\item internal energy
\item magnetization. 
\end{enumerate}
For $\kappa > 0.5 $, the ordered phase is ``antiphase'' and it is difficult
to identify the order parameter unambiguously. In this case, we have measured 
\begin{enumerate}
\item internal energy,
\item layer magnetization (magnetization perpendicular to the direction of
frustration) 
\item length of domain walls running along the direction of
frustration 
\item number of domain-intercepts that are of length 2 over a
straight line along the direction of frustration 
\item number of dislocations measured as the number
of domain walls that do not touch the upper and/or lower boundary.
\end{enumerate}
We shall discuss later (Sec. III) the significance of these quantities in the
context of our work. 

From the measurement of a suitable physical quantity $Q(t)$ at a time $t$, 
the critical point (or, for that 
matter, the critical region) could be identified from the general principle that
at the critical point the quantity $Q(t) - Q(\infty)$ is expected to vanish 
algebraically as a function of time $t$. While this characteristic is handy 
for the case $Q(\infty) =0$, it is not usable when $Q(\infty)  \ne 0$, 
as very large time simulation is required to measure the quantity $Q(\infty)$ 
itself.
In such cases, we have utilized two essential features of critical phenomena :
(i) For a finite lattice, the correlation time will diverge as \cite{binder}
\[ \tau \sim L^z.\]
Hence, the quantity $Q(t)$ will depend strongly on the
system size only at the critical temperature $T_c$ even at finite values of
$t$.
(ii) the equilibrium value $Q(\infty)$ will undergo a sudden change, which 
is detectable even for small size and 
becomes more and more drastic as the system size increases. 

As mentioned above, the results from non-equilibrium relaxation study 
\cite{jap} and the density matrix renormalization group analysis \cite{jap2} 
contradicts the previous studies \cite{bkc_book,jap,sato} as regards the 
extent of the floating phase. This paper confirms the conclusion of the former
two studies by Monte Carlo simulation. One should note that all these
studies agree at sufficiently low temperatures. We measure some observables 
that play a crucial role in the underlying physics and that have not been 
analyzed till now. 

\begin{figure}
\noindent \includegraphics[width = 6cm,angle = 270]{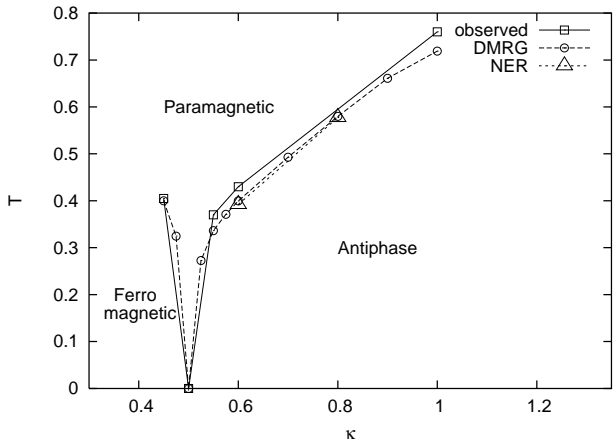}
\caption{\label{fig:Realphase}The phase diagram of the two-dimensional ANNNI 
model as obtained from the present study, from non-linear relaxation (NER)
\cite{jap} and density matrix renormalization group (DMRG) \cite{jap2}.
The temperature is measured in units of $T_c^{(0)} = 0.44069$ as mentioned in 
the text.}
\end{figure}

In Sections II and III we shall present the simulation studies for $\kappa 
< 0.5$ and $\kappa > 0.5$ respectively. All the simulation studies were 
performed with sequential Metropolis algorithm using periodic boundary 
conditions in X and Y directions and the results were averaged over 10 to 50 
realizations. In Section IV we shall study
the correspondence between the 2D ANNNI model and the transverse ANNNI chain
and in Section V present conclusions. Our final conclusion is that
the divergent correlation time exists only along a line.
The phase diagram obtained is presented in Fig.~\ref{fig:Realphase}. 
Everywhere in this communication temperature is measured in unit of 
$T_c^{(0)} = 0.44069$, the critical temperature for nearest-neighbor 
interaction ($\kappa =0$).
This diagram is in qualitative agreement with that obtained by  
Shirahata and Nakamura \cite{jap} and Derian, Gendiar and Nishino \cite{jap2}.
The small difference between our results and those obtained by these authors
seems to be due to the small ($\sim 1000\times 1000$) size of our simulation, 
compared to that of Shirahata and Nakamura ($\sim 6399 \times 6400$).

\section{Simulation Studies for $\kappa < 0.5$}  

\subsection{Energy}

Before we consider the measurements on the 2D ANNNI model itself, let us
break off for a discussion on critical behavior of energy relaxation in 
general. Internal energy $E$ is always unambiguously defined, in contrast to the
order parameter, which for some phase transitions (like our system for 
$\kappa > 0.5$) may not be easy to identify
and measure. However, since $E(\infty) \ne 0$, it is difficult to study 
the time variation of the quantity $E(t) - E(\infty)$, as mentioned above. 
The relaxation of energy has also been studied elsewhere \cite{bray,bunker}. 

\begin{figure}
\noindent \includegraphics[clip,width= 6cm, angle = 270]{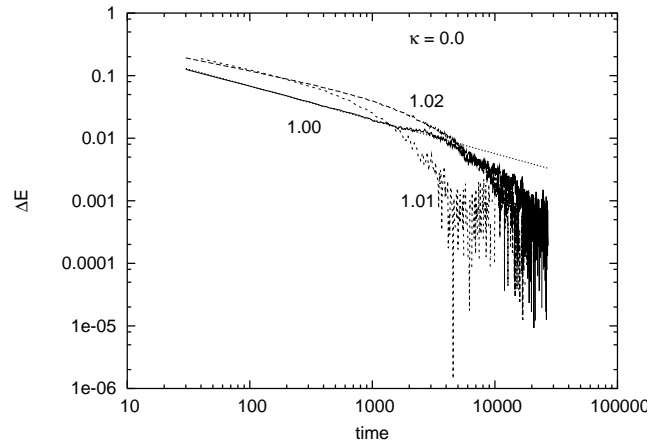}
\caption{\label{fig:kappa00.1000}Energy relaxation for a $1000 \times 1000$ 
lattice for only nearest neighbor interaction i.e. $\kappa = 0$, starting from an ordered 
(all up) configuration. The straight line is a guide to the eye and fits to
$0.82 t^{-0.54}$. }
\end{figure}

We shall now consider the case of $\kappa=0$ (only nearest-neighbor 
interaction), and obtain the exponent,
following standard scaling arguments \cite{ma,bray,janssen}. Starting from the
standard diffusion equation for the (non-conserved) order parameter
$\psi$
\[
\frac{\partial \psi}{\partial t} = - \Gamma \frac{\partial F}{\partial \psi}
\]
one obtains
\[
\frac{\partial F}{\partial t} = - \frac{1}{\Gamma}
\left(\frac{\partial \psi}{\partial t}\right)^2,
\]
where $F$ is the appropriate free energy and $\Gamma$ is a parameter of the 
model. The identity
$E = \partial(\beta F)/\partial \beta$, where $E$ is the total internal energy
then gives,
\be
\frac{\partial E}{\partial t} = - \frac{1}{\Gamma}\left(1 +
\beta \frac{\partial}{\partial \beta}\right)
\left(\frac{\partial \psi}{\partial t}\right)^2.
\ee
The scaling relation for the order parameter may be written as 
\cite{janssen},
\be
\psi(t, \epsilon, L, \psi_0) = b^{-\beta/\nu} \psi(b^{-z}t, b^{1/\nu} \epsilon,
b^{-1}L)
\ee
where $\epsilon=(1-T)$ ($T$ being the temperature measured in units of 
$T_c^{(0)}$), L is the linear dimension of the system, $b$ is the
scaling factor, and $\psi_0$ is the initial value of the order parameter.
Also, $\beta$ and $\nu$ are the static critical exponents, and $z$ is the 
dynamic critical exponent. Choosing $b=t^{1/z}$ and suppressing 
the unimportant arguments $L$ ( $\rightarrow \infty$) and $\psi_0$, we obtain,
\be \psi(t, \epsilon) = t^{-\beta/\nu z} \left[ 
\psi(1,0) + t^{1/\nu z} \epsilon \psi^{\prime}(1,0) \right] \ee
where $\psi^{\prime}$ is the derivative of $\psi$ with respect to $\epsilon$.
Substituting this form in Eq. (2), we have, for large time,
\be
\Delta E \sim t^{-\sigma}
\ee
where $\Delta E$ is the energy difference $E(t) - E(\infty)$ and
$\sigma= 1 - (1 - 2\beta)/\nu z = 0.66$. Our simulations confirm this value
of sigma (Fig.~\ref{fig:kappa00.1000}). 

\begin{figure}
\noindent \includegraphics[clip,width= 6cm, angle = 270]{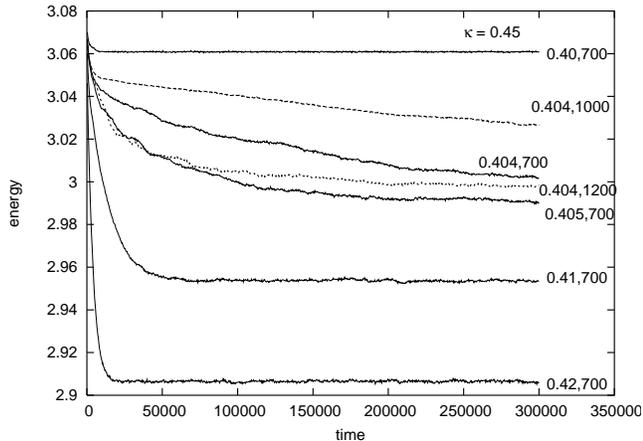}
\caption{\label{fig:relaxk45}
Energy relaxation for $\kappa=0.45$ at temperature $T$ for
$L \times L$ square lattice. The numbers at the right margin indicate $T$ and
$L$ values. It is important to note that for $T=$0.40, 0.405, 0.41 and 0.42,
the curves for $L=$1000 and 1200 coincide with that of $L=700$.}
\end{figure}

\begin{figure}
\noindent \includegraphics[clip,width= 6cm, angle = 270]{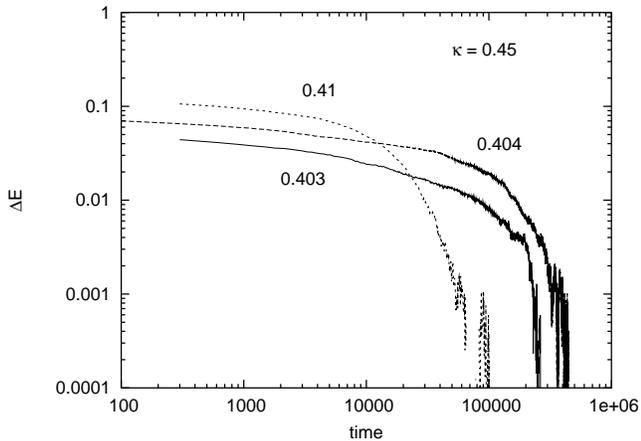}
\caption{\label{fig:kappa45.700}
Energy relaxation for $\kappa=0.45$ for $700 \times 700$ lattice. The numbers
indicate temperature. Note that the linear (algebraic) region is most prominent
for $T=0.404$. The algebraic region fits to $0.17t^{-0.15}$. We could not 
furnish the curve for $T=0.405$ as the system takes too long time to 
equilibrate.}
\end{figure}

In the case of ANNNI model ($\kappa$ non-zero but $< 0.5$) we study the 
energy relaxation starting from ferromagnetic state. Our results for 
$\kappa = 0.45$ is as follows. The $E(t)$ vs $t$ curve for an
$L \times L$ lattice is the same for $L=$ 700, 1000 and 1200 at a temperature 
$T=$0.40 (in units of $T_c^{(0)}$). This happens also for $T=$0.41 but not 
for $T=$0.404, where 
the relaxation behavior depends on $L$ to the largest extent 
(Fig.~\ref{fig:relaxk45}). At $L=$ 700, we could evaluate $E(\infty)$ and 
found that $E(t)-E(\infty)$ vs $t$ plot shows an algebraic decay around 
$T$=0.40 but the 
algebraic region is most extended for $T=$ 0.404 (Fig.~\ref{fig:kappa45.700}).
It is hence concluded that at $\kappa=0.45$, the critical point is $T_c=0.404 \pm 0.002$.
We could not however evaluate $E(\infty)$ for $L=1000$ or 1200 as they 
involve too much computational time. Moreover, even at $T=0.404$,
there is some anomaly in the sense that the decay of $E(t)$ at $L=1200$
is {\em faster} (rather than slower) than that for $L=700$ and 1000 
(Fig.~\ref{fig:relaxk45}). 
A simulation over a longer time scale might resolve the anomaly.

It is interesting to note that at $\kappa=0.45$, $T=0.404$, the exponent 
$\sigma$ has a value $0.15 \pm 0.05$. Although it is a
computationally intensive job to determine $\sigma$ accurately, we
attempted to study the apparent variation of $\sigma$ with $\kappa$ and found
that it does not change much till $\kappa=0.4$ and afterwards decrease markedly.

\subsection{Magnetization}

For $\kappa < 0.5$, the order parameter is magnetisation $M$, whose equilibrium
value is zero at the critical point. Hence, an easy method of locating the
critical point is to investigate where the magnetisation relaxes
algebraically. It is well-known (Eq.(4)) that at $T_c$ the magnetisation
decays as 
\be M \sim t^{-\sigma^{\prime}} \ee
where $\sigma^{\prime} = \beta/\nu z$. For $\kappa=0$, the value of 
$\sigma^{\prime}$ is 0.05734. 

\begin{figure}
\noindent \includegraphics[clip,width= 6cm, angle = 270]{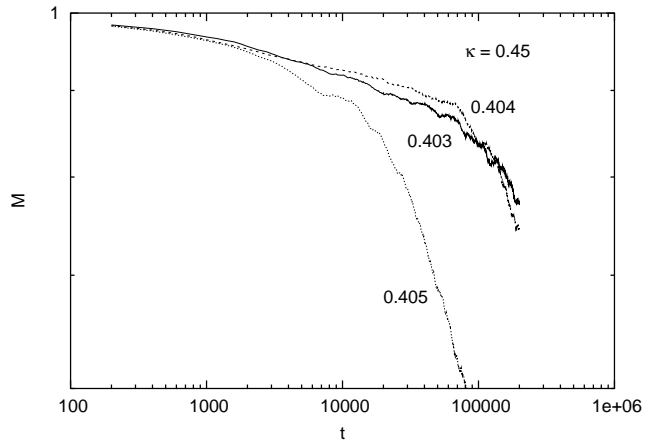}
\caption{\label{fig:relaxmag2.k45}Relaxation of magnetization at $\kappa = 
0.45$ at for $1000 \times 1000$ lattice. The linear (algebraic) region is most 
prominent for $T=$ 0.404 and fits to $1.08t^{-0.017}$.}
\end{figure}

\begin{figure}
\noindent \includegraphics[clip,width= 6cm, angle = 270]{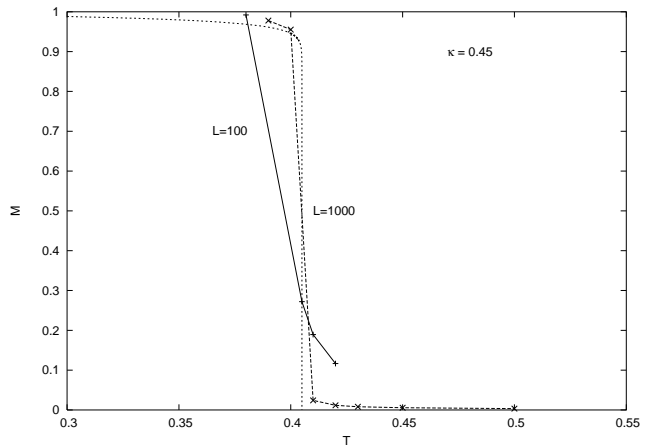}
\caption{\label{fig:mag2.k45}Equilibrium value of magnetization at
$\kappa = 0.45$ as a function of temperature.}
\end{figure}

At $\kappa=0.45$ the magnetization is found to relax 
algebraically only around $T=0.405 \pm 0.002$, which therefore is the critical 
temperature (Fig.~\ref{fig:relaxmag2.k45}).
That the critical temperature at $\kappa=0.45$ lies between 0.40 and 0.41 is 
also verified by the fact that there is a sudden change in the equilibrium 
value of magnetization as temperature increases from 0.40 to 0.41, and that 
this change becomes more and more sudden as the lattice size increases 
(Fig.~\ref{fig:mag2.k45}).

The exponent for magnetization decay turns out to be $\sigma^{\prime} = 
0.02 \pm 0.005$ at $\kappa=0.45$, $T=0.404$. As for the case of energy
relaxation, it is a computationally intensive job to determine 
$\sigma^{\prime}$ accurately. Approximate measurements indicate that 
just like $\sigma$, the exponent $\sigma^{\prime}$ also remains more or
less the same up to $\kappa=0.4$ and starts decreasing markedly at higher 
$\kappa$. Further investigations on the apparent variation of $\sigma$ and
$\sigma^{\prime}$ is in progress.

\section{Simulation Studies for $\kappa > 0.5$}  

In this case, it is difficult to identify the order parameter, as mentioned 
above. It can be easily proved \cite{selke,yeomans,liebman} that, at low 
temperature the system is in a perfectly ordered state with like spins along 
the Y axis and ++~-~- 
pattern repeated along the X direction. The domain walls thus run exactly
parallel to the Y axis. Moving along the X (Y) axis, one finds domains of 
length 2 ($L$), for an $L \times L$ system. As one increases the temperature,
after some temperature $T_c$ (called the lower critical temperature), domains 
of length larger than 2 appear along the X direction. The domain walls now run
not always parallel to the Y axis. They start from the lower boundary and 
terminate at the upper, but they often take small steps parallel to the X axis.
Along the Y axis the domains are now sometimes less than $L$ in length. 
Villain and 
Bak \cite{bak} pointed out that the number of domains that do not touch
{\em both} boundaries is crucial and represents some sort of ``dislocation''. 
This number is almost zero immediately above $T_c$
but suddenly increases at some temperature $T_2$. While $T_c$ marks a second 
order commensurate-incommensurate (Pokrovsky-Talapov type) transition, 
$T_2$ marks a 
Kosterlitz-Thouless type transition. For $T > T_2$ domains of small size appear
in X and Y direction and the system is in a paramagnetic state. For 
$T_c < T < T_2$, Villain and Bak claimed that, the wavelength of modulation 
changes continuously with temperature and the correlation length is infinity, 
leading to a spin-spin correlation that decays algebraically with distance.

To study the two transitions, one at $T=T_c$ and the other at $T=T_2$, we 
measure several quantities always starting the simulation with antiphase as the
initial configuration. 

(1) Internal energy is studied in the same way as done for $\kappa < 0.5$. 

(2) Layer magnetization is defined (following 
\cite{jap}) as the magnetization of a chain along the Y axis averaged 
over all such chains :
\be m_l = \frac{1}{L} \sum^L_{x=1} \mid m_x \mid \ee
where,
\[ m_x = \frac{1}{L} \sum^L_{y=1}  s_{x,y}  \]
Obviously, this quantity should be 1 for $T < T_c$ and zero above $T_2$. 
Hence, the size dependence of the relaxation of $m_l$ and the algebraic 
nature of relaxation of $m_l$ would indicate a diverging correlation length. 
Shirahata and 
Nakamura \cite{jap} have studied this quantity to identify the upper transition
temperature $T_2$ at $\kappa=0.60$ and $0.80$. 

(3) The sum of length of the segments of domain
walls that are parallel to the X axis, divided by the system size
gives a quantity (say, $d_x$) which is strictly zero in the
perfect antiphase, but increases suddenly to some non-zero value at $T_c$.
Its measurement leads to an estimate of $T_c$.

(4) Moving in the X-direction, one may note the lengths of domain intercepts 
encountered, and compute the ratio
\[ f_2 = n_2/n_t \]
where $n_2$ is the number of domains of length 2 and $n_t$ is the total number
of domains. This ratio is 1 for $T < T_c$ and decreases for higher $T$. We 
note the region over which the relaxation of $f_2$ depends on size or is
algebraic in nature. 

(5) The spin domains are identified and the fraction ($f_d$) of domains that 
do not touch the upper and/or lower boundary is counted. (Those which miss
any one boundary is counted with weightage 1 and those which miss both the
boundaries are given weightage 2.) This fraction 
measures the number of ``dislocations'' that drives the Kosterlitz-Thouless 
phase transition
at $T_2$. We do not study the relaxation of this quantity for computational
limitations, but obtain from simulation the equilibrium value away from $T_c$.
Such measurement should lead to an estimation of the Kosterlitz-Thouless 
transition temperature $T_2$, if any.

The study of all these quantities leads us to the conclusion that the floating
phase, i.e. the region between $T_c$ and $T_2$ extends, if at all, over a 
temperature range less than 0.02. Thus, the region of diverging correlation
length exists {\em only along a line}, up to the accuracy of this study. Our 
study was
performed at $\kappa=$ 0.55, 0.60, 1.0. 
The resulting phase diagram is presented
in (Fig.~\ref{fig:Realphase}). However, it is interesting to observe that while the width of the
critical region is 0.02 for $\kappa > 0.5$, it is much less, about 0.001, for
$\kappa < 0.5$.

\subsection{Energy}

\begin{figure}
\noindent \includegraphics[clip,width= 6cm, angle = 270]{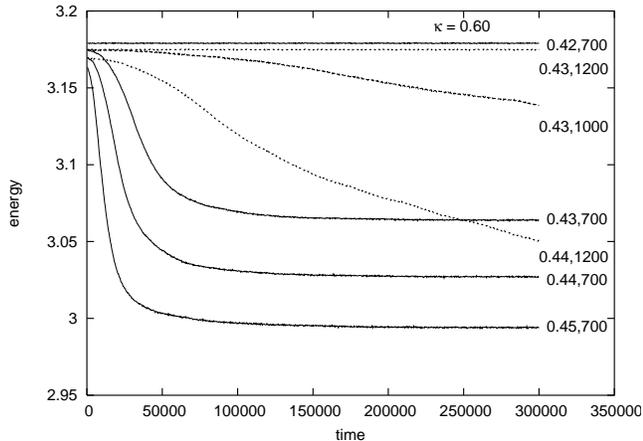}
\caption{\label{fig:relaxk60}
Energy relaxation for $\kappa=0.60$ at temperature $T$ for
$L \times L$ square lattice. The numbers at the right margin indicate $T$ and
$L$ values. For $T=$0.42 and 0.45,
the curves for $L=$1000 and 1200 coincide with that of $L=700$.}
\end{figure}

\begin{figure}
\noindent \includegraphics[clip,width= 6cm, angle = 270]{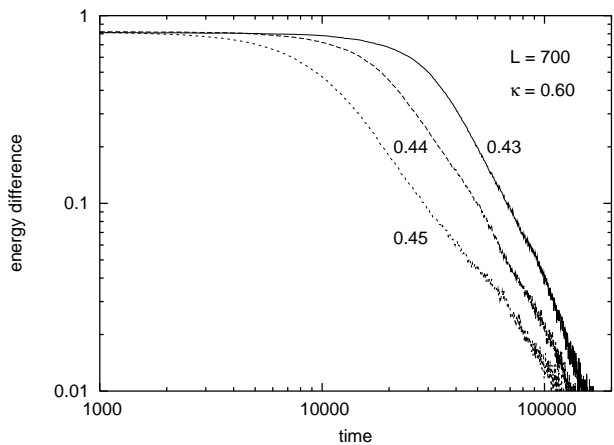}
\caption{\label{fig:kappa60.700}
Energy relaxation for $\kappa=0.60$ for $700 \times 700$ lattice. The numbers
indicate temperature. Note that the linear (algebraic) region is most prominent
for $T=$ 0.44 and fits to $75 \times 10^{6}t^{-1.9}$.}
\end{figure}

The study of energy relaxation for $\kappa > 0.5$ follows closely the procedure
for $\kappa < 0.5$, with the only difference that the initial configuration
is now the antiphase. For $\kappa=0.60$, the energy relaxation depends on 
size predominantly at $T=$ 0.43 and 0.44 (Fig.~\ref{fig:relaxk60}) and 
for $L=$ 700, the energy difference $E(t) - E(\infty)$ shows an algebraic 
decay over an extended region of time at $T = $0.43 and 0.44 
(Fig.~\ref{fig:kappa60.700}). Therefore, at $\kappa=0.60$ 
we identify $T_c$ as $0.44 \pm 0.01$. The curves for $\kappa=0.55$ are qualitatively
similar to that for $\kappa=0.60$, and $T_c$ could be identified as $0.37 \pm 0.01$.
An alternative interpretation of the results could
be that the floating phase would exist, if at all, between $T=$0.43 to 0.44
(0.37 to 0.38) at $\kappa =$ 0.60 (0.55). 
That the energy does not show any critical behavior over an extended range
of temperature, seems to indicate that the floating phase does not exist over
an extended region.

We mention that we could not perform the study of energy relaxation at 
$\kappa=1$, since for
this case a reliable data needs averaging over too many configurations.
 
The exponent for energy relaxation $\sigma^{\prime}$ (see Eq.(5)) is found to 
be $1.7 \pm 0.1$ for $\kappa=$ 0.55 and 0.60. In contrast to the 
findings for $\kappa < 0.5$, we observe no marked variation of
$\sigma^{\prime}$ with $\kappa$.

\subsection{Layer Magnetization}
 
\begin{figure}
\noindent \includegraphics[clip,width= 6cm, angle = 270]{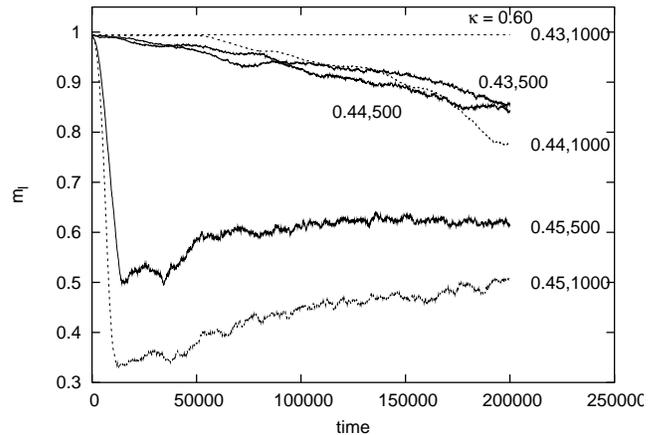}
\caption{\label{fig:relaxlayer_mag1.k60}Relaxation of layer magnetization for $\kappa=0.60$ at temperature 
$T$ for $L \times L$ square lattice. The numbers at the right margin indicate 
$T$ and $L$ values.}
\end{figure}

\begin{figure}
\noindent \includegraphics[clip,width= 6cm, angle = 270]{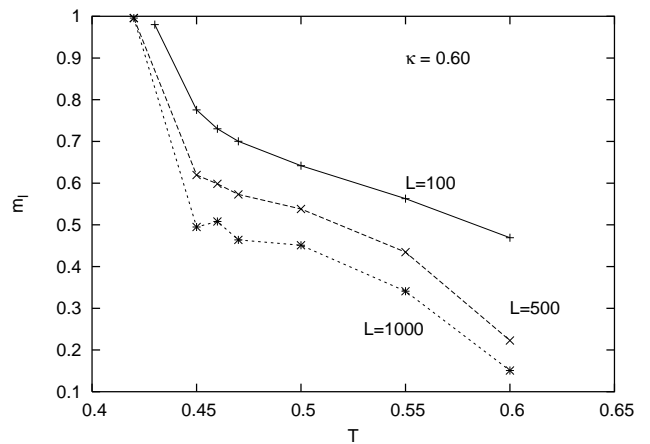}
\caption{\label{fig:layer_mag1.k60}Equilibrium value of layer magnetization at 
$\kappa = 0.60$ as a function of temperature. }
\end{figure}

The relaxation of layer magnetization $(m_l)$ is qualitatively similar for 
$\kappa=$ 0.55, 0.60 and 1.00. At $\kappa=0.60$ the relaxation 
shows critical slowing down and finite size effect at $T= 0.44 \pm 0.01$
(Fig.~\ref{fig:relaxlayer_mag1.k60}), which is therefore the value of $T_2$.
The equilibrium value of layer magnetization also shows a sharp fall (that
becomes sharper as the system size increases) at this temperature, 
at $\kappa= 0.60$ (Fig.~\ref{fig:layer_mag1.k60}).
For $\kappa=0.55$ the value of $T_2$ can be estimated in a similar manner to 
be $0.37 \pm 0.01$.
We could not observe the curve for equilibrium value of layer magnetization
at $\kappa=1$ because for this case one needs too long simulation to
get the equilibrium value. 

\subsection{Length of domain walls parallel to the direction of frustration}

\begin{figure}
\noindent \includegraphics[clip,width= 6cm, angle = 270]{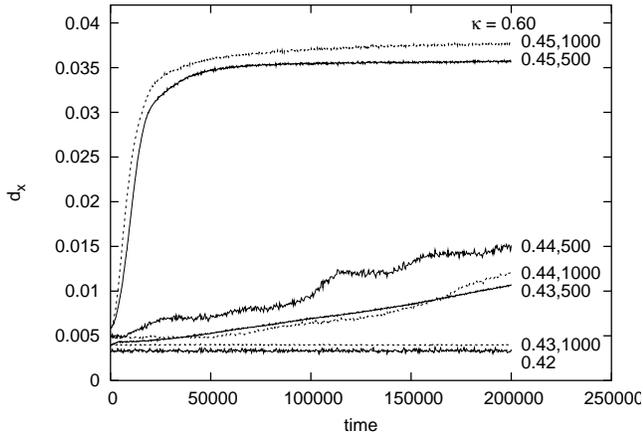}
\caption{\label{fig:relaxmag3.k60}Relaxation of $d_x$ for $\kappa=0.60$ at temperature $T$ for
$L \times L$ square lattice. The numbers at the right margin indicate $T$ and
$L$ values. For $T=$0.42, the curves for $L=$500 and 1000 coincide.}
\end{figure}

\begin{figure}
\noindent \includegraphics[clip,width= 6cm, angle = 270]{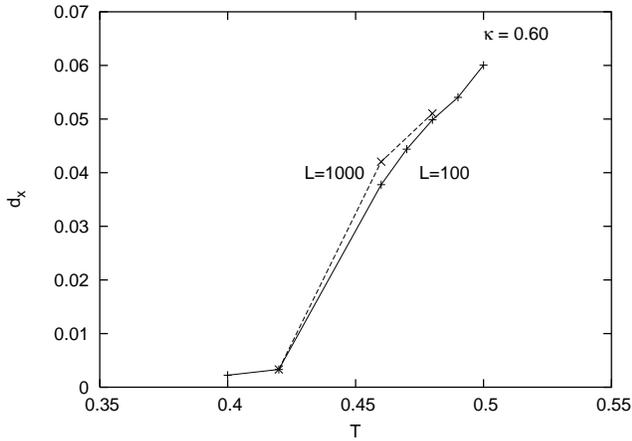}
\caption{\label{fig:mag3.k60}Equilibrium value of $d_x$ at $\kappa = 0.60$ 
as a function of temperature. }
\end{figure}

Like layer magnetisation, this quantity ($d_x$) shows critical slowing down
and finite size effect at $T=$ 0.44 for $\kappa=$ 0.60 
(Fig.~\ref{fig:relaxmag3.k60}). 
The equilibrium value of $d_x$ shows a sharp rise at this temperature
(Fig.~\ref{fig:mag3.k60}), and this rise becomes sharper as the system size 
increases. Hence, 
the study of $d_x$ indicates that $T_c$ is $0.44 \pm 0.01$ for $\kappa=$ 0.60.
In the same manner, the value of $T_c$ is estimated to be $0.37\pm 0.01$  for 
$\kappa=$ 0.55 and $0.76\pm 0.01$  for $\kappa=1.00$.

\subsection{Fraction of domains that have length 2}

\begin{figure}
\noindent \includegraphics[clip,width= 6cm, angle = 270]{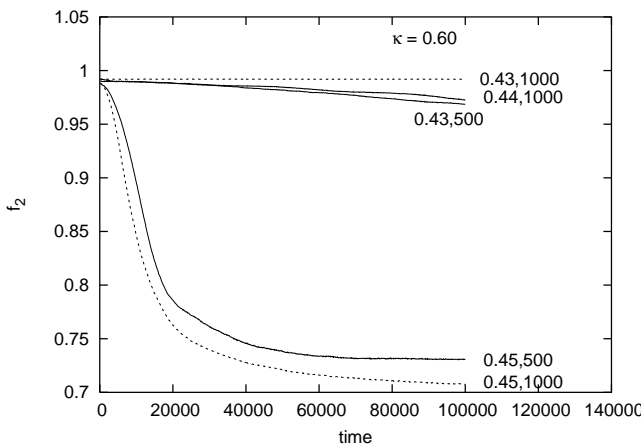}
\caption{\label{fig:relaxdistpair.k60}Relaxation of $f_2$ for $\kappa=0.60$ at temperature $T$ for
$L \times L$ square lattice. The numbers at the right margin indicate $T$ and
$L$ values.}
\end{figure}

\begin{figure}
\noindent \includegraphics[clip,width= 6cm, angle = 270]{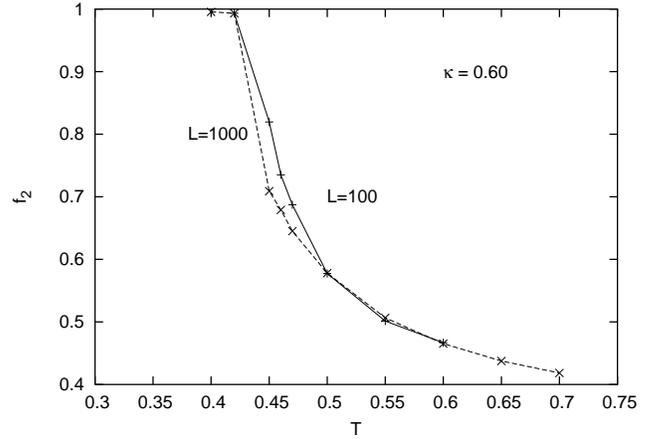}
\caption{\label{fig:distpair.k60}Equilibrium value of $f_2$ at $\kappa = 0.60$ 
as a function of temperature. }
\end{figure}

A study of this quantity ($f_2$) leads to $T_c= 0.44 \pm 0.01$ for $\kappa=0.60$, since 
size-dependent slowing down of the relaxation of $f_2$ is observed at this 
temperature (Fig.~\ref{fig:relaxdistpair.k60}). Moreover, a sharp 
fall of the equilibrium value of $f_2$ is observed at this temperature      
and this fall becomes slightly sharper as the system size increases from
100 to 1000 (Fig.~\ref{fig:distpair.k60}). In a similar manner, the critical 
temperature for $\kappa=0.55$ and 1.00 is obtained as $0.37 \pm 0.01$ and $0.76 \pm 0.01$ 
respectively.

\subsection{Fraction of domains that do not touch the boundary ($f_d$)}

\begin{figure}
\noindent \includegraphics[clip,width= 6cm, angle = 270]{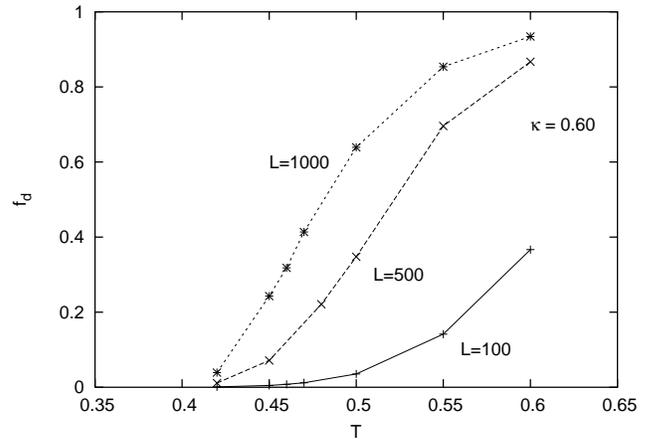}
\caption{\label{fig:cl5_full.k60}Equilibrium value of $f_d$ at $\kappa = 0.60$ 
as a function of temperature.}
\end{figure}

We could not study the relaxation of this quantity since averaging over too
many realisations is necessary for a reasonably smooth curve. Rather, we could
measure the 
equilibrium value at temperatures where the relaxation was not prohibitively 
slow. It is found that for $\kappa = 0.60$, the critical temperature lies
between 0.42 and 0.45, since a sudden change is observed in the equilibrium
value of $f_d$ in this temperature range and that this change becomes more and 
more sudden as the lattice size increases (Fig.~\ref{fig:cl5_full.k60}). 
Similar behaviour is also observed for $\kappa = $ 0.55 between temperature 
0.35 and 0.40. This study could not be done
for $\kappa = $ 1.0 because of computational limitation (one has to average
over a good number of realisations to get reliable data). Inspite of the 
computational difficulties for the study of $f_d$, it is clear that the study
of this quantity excludes the possibility of critical region with width larger
than 0.02 along temperature axis.

\begin{figure}
\noindent \includegraphics[clip,width= 6cm, angle = 0]{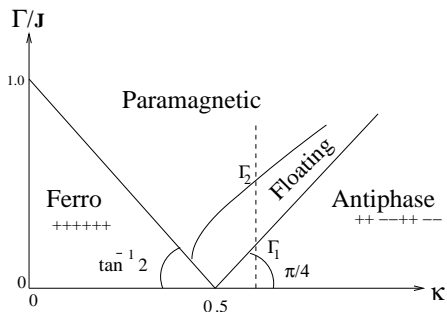}
\caption{\label{fig:modPhase2}Schematic phase diagram for Transverse ANNNI model (with Hamiltonian ${\mathcal H}_q$ of Eq. (8)), after ref. [13]}
\end{figure}

\section{Mapping to quantum model}  

A quantum Ising model in $d$ dimension is related to a classical Ising model
in $d+1$ dimension by Suzuki-Trotter transformation 
\cite{suzuki,bkc_book,arnab} and this relation is the basic idea behind 
quantum Monte Carlo
algorithm. This transformation for the ANNNI model has
been discussed in detail by Arizmendi \cite{ariz} and can be summarised
as follows. The quantum Ising Hamiltonian ${\mathcal H}_q$ for the 
one-dimensional transverse ANNNI model with $N$ sites is given by,
\be {\mathcal H}_q = - J^{\prime} \sum_{j=1}^N (s^z_j s^z_{j+1} -
\kappa s^z_j s^z_{j+2}) - \Gamma \sum_{j=1}^N s^x_j. \ee
The ground state for this model is equivalent to a classical Ising model in 
two dimension with Hamiltonian 
\be {\mathcal H}_{cl} = - \sum_{x=1}^N \sum_{y=1}^{mn} 
J_{q}s_{x,y}[(s_{x+1,y}-\kappa s_{x+2,y}) + p s_{x,y+1}]  \ee
in the limit $(m \rightarrow \infty)$ and $(n \rightarrow \infty)$ at a 
temperature $T_q$ where
\be \frac{k_B T_q}{J_q} = \frac{m\Gamma}{J^{\prime}} \ee 
($k_B$ is the Boltzmann constant) and 
\be p = \frac{m\Gamma}{2J^{\prime}}\log [\coth(1/m)] \ee
Obviously, the nearest neighbour interaction in the $y$ direction is $p$ times
the same in the $x$ direction and as $m \rightarrow \infty$ this ratio $p$
also tends to infinity.

As mentioned earlier, the Hamiltonian ${\mathcal H}_q$ of Eq. (8), describing 
a transverse ANNNI chain, has also been studied widely and 
numerical and (approximate) analytic studies \cite{bkc_book,rieger,amit,CD2} 
had predicted the existence of floating phase over a wide region extending from
$\kappa < 0.5$ to $\kappa > 0.5$ (Fig.~\ref{fig:modPhase2}). Why then this 
extensive presence of floating phase does not occur in the phase diagram of
Fig.~\ref{fig:Realphase}? It seems that the explanation is the 
following.
For the classical Hamiltonian ${\mathcal H}_{cl}$ of Eq. (9),
the strength of nearest-neighbor interaction
in the Y direction is $p$ times stronger than that in the X direction, thus
stabilizing the order in the Y direction. This raises both the upper critical
temperature $T_2$ (at which the order in the Y direction breaks down) and the
the lower critical temperature $T_c$ (at which the order in the X direction
breaks down). Both the temperatures are raised by almost the same amount
(thus the floating phase always has negligible width) but the amount of rise
depends on the value of $p$. Now consider the quantum Hamiltonian 
${\mathcal H}_q$.
For a given $\kappa$ if the floating phase extends from $\Gamma_1$ to
$\Gamma_2$ ($\Gamma_2 > \Gamma_1$, the difference $\Gamma_2 - \Gamma_1$ being 
appreciable), then the value of $p$ (say, $p_2$) corresponding to $\Gamma_2$ 
will be proportionately larger than the value of $p$ (say, $p_1$)
corresponding to $\Gamma_1$ (see Eq. (11)). Although for every value of $p$, 
one has $T_c \approx T_2$,
the common value of $T_c$ and $T_2$ at $p_2$ is appreciably larger than that
at $p_1$. This shows that the {\em presence} of floating phase over a wide
parameter region for the quantum model is consistent with the {\em absence}
of the same for the classical model.  

\begin{table}[h]
\caption{Conclusions regarding the critical region from the study of
various
observables (Sec. II and III here). $T_0$ is the temperature (in units
of $T_c^{(0)} = 0.44069$) at which there is a critical region of width
$< 0.02$}
\vskip 0.5cm
{\large{
\hspace*{0.8cm}
\begin {tabular}{|l|l|}
\hline
\hspace*{1cm}$\kappa$\hspace*{1cm} & \hspace*{1cm}$T_0$\hspace*{1cm} \\
\hline
\hspace*{1cm}0.45\hspace*{1cm} & \hspace*{1cm}0.404\hspace*{1cm} \\
\hspace*{1cm}0.55\hspace*{1cm} & \hspace*{1cm}0.37\hspace*{1cm} \\
\hspace*{1cm}0.60\hspace*{1cm} & \hspace*{1cm}0.44\hspace*{1cm} \\
\hspace*{1cm}1.0\hspace*{1cm}  & \hspace*{1cm}0.76\hspace*{1cm} \\
\hline
\end{tabular}}}
\end{table}

\section{Conclusions}  

(1) For the 2D ANNNI model, diverging correlation length and
algebraically decaying spin-spin correlation exists only along a ``line''.
We present in Table 1 the temperature $T_0$ such that within the temperature 
range $(T_0 - 0.01) < T < (T_0 + 0.01)$ lie both the upper critical 
temperature $T_2$ and the lower critical temperature $T_c$. We did not conduct
the study at temperature intervals smaller than 0.01 except at $\kappa=0.45$.\\
(2) The phase diagram is topologically different for 2D ANNNI model and its
Suzuki-Trotter counterpart, the 1D transverse ANNNI chain. This is in 
contradiction with the fact that often a quantum model and its corresponding 
classical counterpart show topologically similar phase diagram and have the
same critical indices at the transition point. Two examples of such behavior 
are (i) 2D Ising model and 1D transverse Ising model with nearest neighbor
interaction ($\kappa=0$) \cite{mattis} and (ii) XYZ chain and eight vertex 
model \cite{baxter}.

\begin{acknowledgements}
The work of one author (AKC) was supported by UGC fellowship.
We also acknowledge the financial support from DST-FIST for computational 
facility.\\\end{acknowledgements}

\end{document}